\RequirePackage{fix-cm}
\documentclass[twocolumn]{svjour3}          % twocolumn
\smartqed  % flush right qed marks, e.g. at end of proof
\usepackage{graphicx}
\usepackage{setspace}
\usepackage{multirow}
\usepackage{amsmath}
\usepackage{subfigure}
\usepackage{marvosym}
\usepackage{threeparttable}
\usepackage{amssymb}

\begin{document}
\newcommand{\tabincell}[2]{
\begin{tabular}{@{}#1@{}}#2\end{tabular}
}

%\doublespacing

\title{Optimizing production scheduling of steel plate hot rolling for economic load dispatch under time-of-use electricity pricing}

\titlerunning{Optimizing production scheduling of steel plate hot rolling for ELD under TOU pricing}        % if too long for running head

\author{
        Mao Tan         \and
        Hua-li Yang       \and
        Bin Duan         \and
        Yong-xin Su     \and
        Feng He
}


\institute{
        M. Tan(\Letter) \and H.L. Yang \and B. Duan \and Y.X. Su \at
              Hunan Province Cooperative Innovation Center for Wind Power Equipment and Energy Conversion,
              College of Information Engineering, Xiangtan University, Xiangtan, China \\
              \email{mr.tanmao@gmail.com}           %  \\
            \and
           F. He \at
              Department of Energy and Environmental Protection, Xiangtan Iron and Steel Corporation Ltd, Xiangtan, China
}

\date{Received: date / Accepted: date}

\maketitle

\begin{abstract}
Time-of-Use (TOU) electricity pricing provides an opportunity for industrial users to cut electricity costs. Although many methods for Economic Load Dispatch (ELD) under TOU pricing in continuous industrial processing have been proposed, there are still difficulties in batch-type processing since power load units are not directly adjustable and nonlinearly depend on production planning and scheduling. In this paper, for hot rolling, a typical batch-type and energy intensive process in steel industry, a production scheduling optimization model for ELD is proposed under TOU pricing, in which the objective is to minimize electricity costs while considering penalties caused by jumps between adjacent slabs. A NSGA-II based multi-objective production scheduling algorithm is developed to obtain Pareto-optimal solutions, and then TOPSIS based multi-criteria decision-making is performed to recommend an optimal solution to facilitate filed operation. Experimental results and analyses show that the proposed method cuts electricity costs in production, especially in case of allowance for penalty score increase in a certain range. Further analyses show that the proposed method has effect on peak load regulation of power grid.
\end{abstract}

\section{Introduction}
\label{intro}
Time-of-Use (TOU) electricity pricing, a practical demand response program implemented by many power suppliers to
improve the peak load regulation ability of power grid, provides an opportunity for electricity users to implement Economic Load Dispatch (ELD), i.e., cut electricity costs by reducing power loads during on-peak periods and shifting loads from on-peak to off-peak periods.

\begin{figure*}[tbp]
  \includegraphics[width=0.90\textwidth]{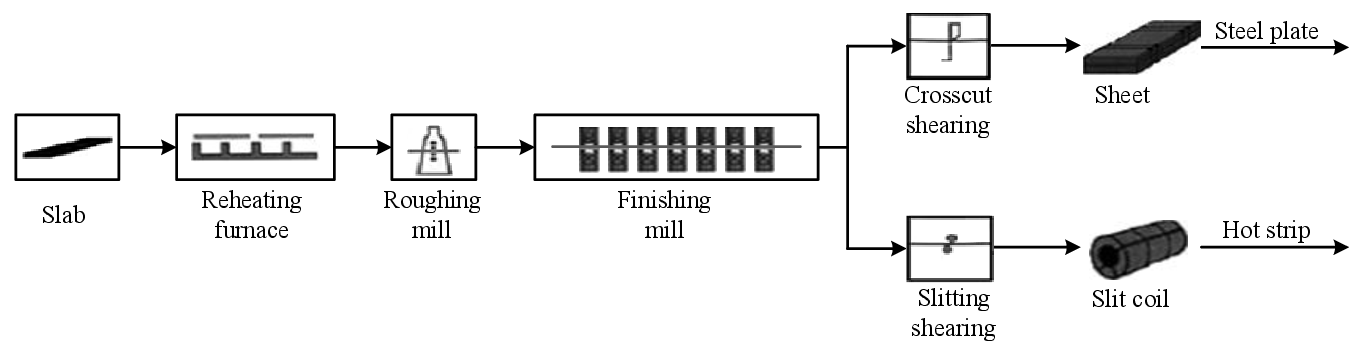}
\caption{A process flow diagram of the hot rolling production procedure}
\label{fig:1}       % Give a unique label
\end{figure*}

Unlike conventional energy conservation to reduce absolute energy consumption, optimizing electricity costs under TOU pricing means that industrial users adjust their production schedule to avoid on-peak time periods, which will have significant effect on cutting electricity costs. In recent years, ELD under TOU pricing has become a hot area. Shrouf et al. \cite{shrouf2014optimizing} proposed a single machine scheduling problem, in which each time period has an associated price and the objective is to minimize electricity costs while considering traditional scheduling performance measures. Fang et al. \cite{fang2014scheduling} also considered job scheduling on a single machine to minimize total electricity costs under TOU pricing and proposed the algorithms for uniform-speed and speed-scalable machine environments respectively. Mitra et al. \cite{mitra2012optimal} formulated a mixed integer linear programming for continuous industrial processing, which allows optimal production planning, and provided a case study for time horizon of one week and hourly changing electricity prices. Furthermore they improved the model with integration of operational and strategic decision-making \cite{mitra2014optimal}. Ashok \cite{ashok2006peak} presented a theoretical model for batch-type load processing and proposed an integer programming method to reschedule their operations to reduce electricity costs under time varying electricity price, but the model is an abstract theoretical model and difficult to be applied to production directly. Wang et al. \cite{wang2012integrated} proposed an optimization model to minimize electricity costs for steel plant, in which both power generation scheduling and batch production scheduling were considered, although the model has been convinced to be effective under TOU pricing, the results can not always be optimal because the production load units are determined by fixed production planning and scheduling.

The above analyses motivates the potential for more benefits by ELD under TOU Pricing in hot rolling production scheduling. Until now, most of the related literatures focused on specific part of the problem or the abstract simplified problem, thus there are still difficulties since the power load units are not directly adjustable and nonlinearly depend on the results of production planning and scheduling.

Hot rolling, a typical batch-type and energy intensive process in steel production with characteristics of strong schedulability, has become an important aspect of production organization and energy saving \cite{bante2013energy}. The general process flow of hot rolling production is illustrated in Figure~\ref{fig:1}. Hot rolling is mainly organized and carried out by batch scheduling program in steel mill, the primary task of which is arranging and sequencing slabs into rolling units to smooth jumps in width, gauge, and hardness between adjacent slabs, all of these will directly affect product quality. Hot rolling production scheduling has attracted attention from academia and industry for a long time. An early method proposed by Kosiba et al. treated steel production scheduling as a discrete event sequencing problem, and thus formulated it as a traveling salesman problem \cite{kosiba1992discrete}. Lopez et al. \cite{lopez1998hot} formulated the problem as a generalized prize collecting traveling salesman problem with multiple conflicting objectives and constraints, and proposed a heuristic tabu search method to determine good approximate solutions. Tang and Wang \cite{tang2006iterated} proposed a modified genetic algorithm based on the multiple travelling salesman problem. Chen et al. \cite{chen2008production} formulated the problem as a nonlinear integer programming model, and later it is corrected by Kim \cite{kim2010some} and changed to a linear programming model. Furthermore, Alidaee and Wang \cite{alidaee2012integer} proposed a corrected integer programming formulation and reduced the quantity of variables. Nevertheless, most of proposed models are single objective or transformed models based on weighted-sum approach. Jia et al. \cite{jia2012decomposition} formulated the problem as a multi-objective vehicle routing problem with double time windows and proposed a decomposition-based hierarchical optimization algorithm to solve it. Soon after, he proposed a P-MMAS algorithm to solve the problem, multi-criteria decision-making is performed to recommend the optimal solution from the Pareto frontier \cite{jia2013multi}. Moon et al. \cite{moon2013optimization} proposed a production scheduling model with time-dependent and machine-dependent electricity cost, in which makespan was considered by using the weighted sum objective but batch sizing was not considered, which is obviously simpler than batch scheduling problem. Because of complexity of batch sizing problem, Sarakhsi et al. \cite{sarakhsi2016anew} proposed a hybrid algorithm of scatter search and Nelder-Mead algorithms to improve the performance of solving algorithm.

Due to high energy consumption and rising energy costs in hot rolling production \cite{puttkammer2014hot}, energy saving has also been considered combined with the traditional objective mentioned above. As is shown in figure~\ref{fig:1}, slabs are heated to high temperature before being rolled, the total energy consumed in heating is affected by batch schedule. Since Direct Hot Charge Rolling (DHCR) has significant benefits on energy cost, great efforts have been made to improve the ratio of DHCR while performing batch scheduling \cite{zhao2009two,shan2013research}. Besides that, optimization of rolling schedule by adjusting thickness reduction ratio of slabs between the rolling passes, another way to reduce power consumption that used to drive rolling motor, has also been proposed \cite{tan2014energy,yang2008genetic,li2013multi}.

As mentioned previously, most methods of hot rolling production scheduling concentrate on internal production organization. Although some technical means have been proposed and applied to achieve energy conservation, their potential would be exhausted due to equipment and technology constraints. In this context, methods utilizing favorable external environments should be explored for energy saving. TOU pricing provides an opportunity to reduce electricity costs, but until now papers to implement ELD under TOU pricing for hot rolling production are few published.

This paper considers the Hot Rolling Production Scheduling Problem (HRPSP) as a mixture of batch scheduling problem and time-dependent job-shop scheduling problem. The rolling units, modeled as power load units, are planned and scheduled according to TOU prices. Primary objective of the proposed model is to minimize electricity costs while considering the traditional objective to minimize penalties caused by jumps between adjacent slabs. A multi-objective optimization model and corresponding solving algorithm are additionally proposed.

The rest of this paper is classified as follows: in Section \ref{sec:2}, characteristics of the problem and opportunities under TOU pricing are presented, and a mathematical model with objective to minimize electricity costs in production is formulated. A multi-objective optimization algorithm is developed in Section \ref{sec:3} to solve the problem. Section \ref{sec:4} is dedicated to the experimental procedure and results to evaluate the proposed method, also the peak load regulation effect and robustness of the proposed method is further discussed. Finally, conclusion and future research planning are given in Section \ref{sec:5}.
\begin{figure*}[tbp]
\begin{center}
  \includegraphics[width=0.9\textwidth]{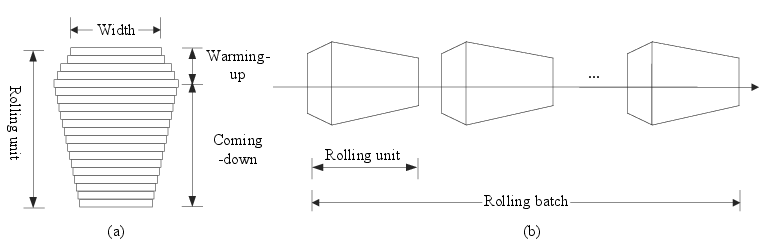}
\caption{Diagrammatic sketch of batch scheduling: (a) rolling unit, (b) rolling batch.}
\label{fig:2}       % Give a unique label
\end{center}
\end{figure*}
\section{Problem description and formulation}
\label{sec:2}
 HRPSP is an extremely complex problem which has significant influence on product quality, production efficiency and energy consumption. In this paper, we study the Hot Rolling Batch Scheduling Problem (HRBSP) combined with the Job-shop Scheduling Problem (JSP), where HRBSP focuses on how rolling units be organized and the JSP concentrates on when the rolling units be processed.
\subsection{Problem description}
\label{sec:2.1}
Hot rolling batch scheduling is a key process in hot rolling. The task of HRBSP, as is depicted in Figure~\ref{fig:2}, is to select, group, and sequence slabs into rolling units with the constraints of production capacity and rolling rules. Each rolling unit has a coffin-shaped width profile consisting of a warming-up section and a coming-down section. In the previous section slabs are arranged from narrow to wide to warm up the rolls, and in the later section slabs are scheduled with decreasing width to avoid marking the coils surface. The major part of a rolling unit is the coming-down section, in which the quality of rolling mainly depends on the sequence of slabs. In most cases, the warming-up section is trivial and can be determined manually.

Several constraints restrict the scheduling, the most important one of which is to smooth jumps in width, gauge and hardness between adjacent slabs. Other constraints, such as cumulative rolling length of slabs in a rolling unit, continuous rolling length of slabs with same width, etc., are also considered to ensure product quality and production capability.

Because hot rolling is a key energy intensive process in steel industry, many approaches, such as optimization of batch scheduling with the objective of improving DHCR ratio and optimization of reduction schedule, have been proposed to achieve energy saving. In smart grid, TOU electricity pricing, which is one of the most commonly implemented demand response programs \cite{palensky2011demand}, provides a new opportunity for steel mill to achieve ELD in hot rolling production, which means cutting costs by shifting loads according to the electricity price.

\begin{figure*}[tbp]
\begin{center}
  \includegraphics[width=0.9\textwidth]{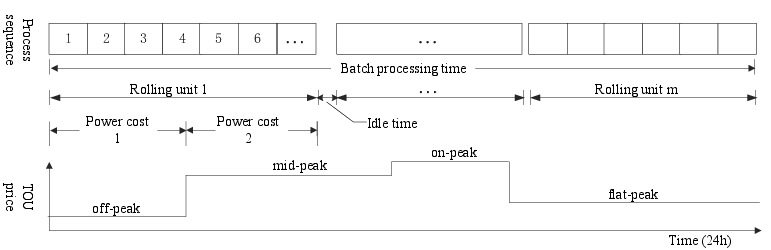}
\caption{Relationship between production scheduling and electricity costs under TOU pricing}
\label{fig:3}       % Give a unique label
\end{center}
\end{figure*}

As is shown in Figure~\ref{fig:3}, a whole day is partitioned into four types of periods based on the price of electricity: on-peak, mid-peak, flat-peak and off-peak period. We can see that the power cost for each rolling unit, which is not only determined by the quantity of power demand but also dependent on the corresponding electricity pricing, should be accumulated piecewise during the processing time.

Compared with flat electricity pricing, the objective of ELD under TOU pricing is to minimize total power cost, including charges for power consumed from shifting loads. In this paper, we assume that rolling units can be scheduled freely, therefore no operating costs from load shifting are included. Consequently, rolling production is encouraged during off-peak periods and discouraged during on-peak periods. In addition, we should know that the scheduling on fixed jobs are not always optimal, so the scheduled jobs, which means the rolling units obtained by hot rolling batch scheduling, should be created and associated to their operation time. Finally, the problem is turned into optimal scheduling for minimizing the electricity costs that determined by batch scheduling solution and job-shop scheduling solution under specified electricity pricing, while the traditional objective that smoothing changes between adjacent slabs should not be ignored to ensure product quality.

\subsection{Mathematical formulation}
\label{sec:2.2}
We interpret the basic model of the HRBSP as a Vehicle Routing Problem (VRP), which is a classical combinatorial optimization problem. In the model, it can be considered that each rolling unit is a vehicle within limited capacity and each slab is a customer that should be visited at most once. Suppose that there are $n$ slabs to be scheduled into $m$ rolling units, the objective of the problem is to determine $m$ routes (rolling units) to minimize the total distance traveled (penalties caused by jumps between adjacent slabs).

The variables used in formulation are listed as follows.
\begin{itemize}
\item [$N$]--a set of slabs, $N=\{1,2,\ldots,n\}$;
\item [$M$]--a set of rolling units, $M=\{1,2,\ldots,m\}$;
\item [$T$]--a set of time periods, $T=\{1,2,\ldots,t\}$;
\item [$\pi_j$]--electricity price during time period $j$;
\item [$W_i$]--power demand of slab $i$ during rolling procedure;
\item [$l_j$]--rolling length of slab $j$;
\item [$p_i$]--processing time for slab $i$;
\item [$P_{ij}$]--the penalty for rolling slab $j$ immediately after slab $i$, where $P_{ij}=p_{ij}^w+p_{ij}^g+p_{ij}^h$, $p_{ij}^w$, $p_{ij}^g$ and $p_{ij}^h$ respectively represent the contribution due to width, gauge, and hardness;
\item [$s_{ij}$]--binary variable with value 1 if the widths of slab $i$ and $j$ are same, otherwise 0;
\item [$ts_i$]--processing start time of slab $i$;
\item [$L$]--lower bound of the cumulative length of slabs that scheduled in a single rolling unit;
\item [$U$]--upper bound of the cumulative length of slabs that scheduled in a single rolling unit;
\item [$R$]--upper bound of the cumulative length of slabs with same width in a single rolling unit;
\item [$TS$]--total time that can be allocated for production;
%\item [$d_{i}^{j}$]--binary decision expression to identify which time period slab $i$ is processed in time periods $j$.
\end{itemize}

Five decision expressions are defined to identify the scheduling solution as follows.
\begin{displaymath}
x_{ij}^{k} = \left\{ \begin{array}{ll}
1 & \textrm{if slab $j$ is immediately after slab $i$ in unit $k$,}\\
0 & \textrm{otherwise.}
\end{array} \right.
\end{displaymath}
\begin{displaymath}
y_{i}^{k} = \left\{ \begin{array}{ll}
1 & \textrm{if slab $i$ is scheduled in rolling unit $k$,}\\
0 & \textrm{otherwise.}
\end{array} \right.
\end{displaymath}
\begin{displaymath}
r_{ij}^{k} = \left\{ \begin{array}{ll}
1 & \textrm{if slab $j$ is rolled after slab $i$ in rolling unit $k$,}\\
0 & \textrm{otherwise.}
\end{array} \right.
\end{displaymath}
\begin{displaymath}
d_{i}^{j} = \left\{ \begin{array}{ll}
1 & \textrm{if slab $i$ is processed in time periods $j$,}\\
0 & \textrm{otherwise.}
\end{array} \right.
\end{displaymath}

{$v_i$}, a positive integer or 0, is a variable to indicate the idle time allocated to rolling unit $i$ before production.

Note that production efficiency may not always be the only one target in engineering, especially in condition of production capacity is abundant, then the target of our model is to minimize electricity costs on the premise of processing all products in given time horizon.
According to basic VRP model combined with consideration of relationship between slab processing sequence and processing time as shown in Figure~\ref{fig:3}, we formulate the hot rolling production optimization problem as
\begin{spacing}{1.1}
\begin{equation}\label{eq:2-1}
  \min f_1=\sum_{k\in M}\sum_{i\in N}\sum_{j\in N}P_{ij}\cdot x_{ij}^k
\end{equation}
\begin{equation}\label{eq:2-2}
  \min f_2=\sum_{j\in T}\left(\pi_j\cdot\sum_{i\in N}W_i\cdot d_{i}^j\right)
\end{equation}
\end{spacing}
%\begin{spacing}{1.5}
%\end{spacing}
s.t.
\begin{spacing}{1.1}
\begin{equation}\label{eq:2-3}
  \sum_{i\in N}x_{ij}^k=y_j^k, j\in N, k\in M
\end{equation}
\begin{equation}\label{eq:2-4}
  \sum_{j\in N}x_{ij}^k=y_i^k, i\in N, k\in M
\end{equation}
\begin{equation}\label{eq:2-5}
  \sum_{k\in M}y_{i}^k=1, i\in N
\end{equation}
\begin{equation}\label{eq:2-6}
  \sum_{i\in N}r_{ij}^k\cdot s_{ij}\cdot l_j\leq R, j\in N, k\in M
\end{equation}
\begin{equation}\label{eq:2-7}
  L\leq\sum_{i\in N}y_{i}^k\cdot l_i\leq U, k\in M
\end{equation}
\begin{equation}\label{eq:2-8}
  0 \leq \sum_{i\in M}v_i\leq TS-\sum_{i\in N}p_i
\end{equation}
\begin{equation}\label{eq:2-9}
  \sum_{k\in M}r_{ij}^k\leq 1, i\in N,j\in N
\end{equation}
\begin{equation}\label{eq:2-10}
  x_{ij}^k\leq r_{ij}^k, i\in N,j\in N,k\in M
\end{equation}
\begin{equation}\label{eq:2-11}
  r_{ij}^k\leq y_i^k, i\in N,j\in N,k\in M
\end{equation}
\begin{equation}\label{eq:2-12}
  r_{ij}^k\leq y_j^k, i\in N,j\in N,k\in M
\end{equation}
\end{spacing}
%\begin{spacing}{1}
%\end{spacing}
where objective $f_1$ is the traditional objective to ensure product quality, which means to minimize the total penalties caused by jumps between adjacent slabs, and objective $f_2$ means to minimize the total electricity costs in hot rolling production, in which $d_{i}^{j}$ can be further formulated as
\begin{equation}\label{eq:2-13}
d_{i}^{j} = \left\{ \begin{array}{ll}
1 & \textrm{if } \sum_{\alpha<j}{\lambda_\alpha}\leq ts_i<\sum_{\alpha\leq j}\lambda_\alpha, \\
0 & \textrm{otherwise.}
\end{array} \right.
\end{equation}
where the condition correspond to $d_i^j=1$ means that slab $i$ is processed in time period $j$.
Note that variable $ts_i$ is not only determined by which rolling units the slab is scheduled in, but also depended on the processing time of previous slabs and the idle time allocated for rolling units, then it can be expressed as
\begin{equation}\label{eq:2-14}
ts_i=\sum_{\delta \in M}y_i^\delta\cdot(\sum_{\beta < \delta}\sum_{\alpha \in N}y_\alpha^\beta\cdot p_\alpha +\sum_{\alpha \leq \delta}v_\alpha +\sum_{\beta=\delta}\sum_{\alpha \in N}r_{\alpha i}^\beta\cdot p_\alpha)
\end{equation}
where $\delta$ is a traversal variable to search the rolling unit that slab $i$ is allocated in, expression in brackets means the cumulative time before processing slab $i$. If slab $i$ is not allocated in rolling unit $\delta$ , the expression in brackets would be ignored because $y_i^\delta=0$.

Constraints (\ref{eq:2-3}) and (\ref{eq:2-4}) specify the sequence of slabs in a rolling unit.
Constraint (\ref{eq:2-5}) ensures that each slab can be scheduled only once.
Constraints (\ref{eq:2-6}) restrict the cumulative length of continuously rolled slabs with same width in each rolling unit.
Constraint (\ref{eq:2-7}) means rolling mill production capacity, which restricts the lower and upper bounds of cumulative length of slabs in each rolling unit.
Constraint (\ref{eq:2-8}) means that the total idle time allocated for rolling units can't be greater than margin of production capability.
Constraints (\ref{eq:2-9})--(\ref{eq:2-12}) restrict the value of $r_{ij}^k$, $x_{ij}^k$ and $y_i^k$ according to their logical relationship.

\section{Production scheduling optimization method}
\label{sec:3}
As known that VRP is a classical NP-hard problem, it is hard to find the optimal solution for large scale problem. Since there are a large number of slabs in the day-ahead scheduling problem combined with complex objective functions, such as $f_2$ with quadratic equation (\ref{eq:2-14}), it is difficult to find the exact optimal solution, even a feasible solution.
In this paper, the production scheduling method consists of two stages. In the first stage, objectives shown in Eq. (\ref{eq:2-1})-(\ref{eq:2-2}) are optimized simultaneously, and a set of Pareto-optimal solutions is generated by the multi-objective optimization algorithm. In the second stage, a TOPSIS based multi-criteria decision-making is performed to recommend an optimal solution to facilitate field operation.

\subsection{NSGA-II based multi-objective optimization}
Recently, many swarm intelligence algorithms are introduced to solve complex optimization problem, in which Non-dominated Sorting Genetic Algorithm with Elitism (NSGA-II) that proposed by Deb \cite{deb2000fast} is a typical method to solve multi-objective problem. NSGA-II has been widely used to solve combinatorial optimization problems in engineering, such as hydro-thermal power scheduling problem \cite{deb2007dynamic}, job sequencing problem \cite{bandyopadhyay2012proposing} and flow-shop scheduling problem \cite{asefi2014hybrid}.
In this paper, a NSGA-II based Multi-objective Production Scheduling Algorithm (MOPSA) is developed to solve the HRPSP, some personalized changes are made to instantiate the algorithm, in which the most important things are designing customized chromosome code and genetic operators to adapt specific problem.

\subsubsection{Chromosome encoding}
\label{sec:3.1}

\begin{figure}[bp]
\begin{center}
  \includegraphics[width=0.5\textwidth]{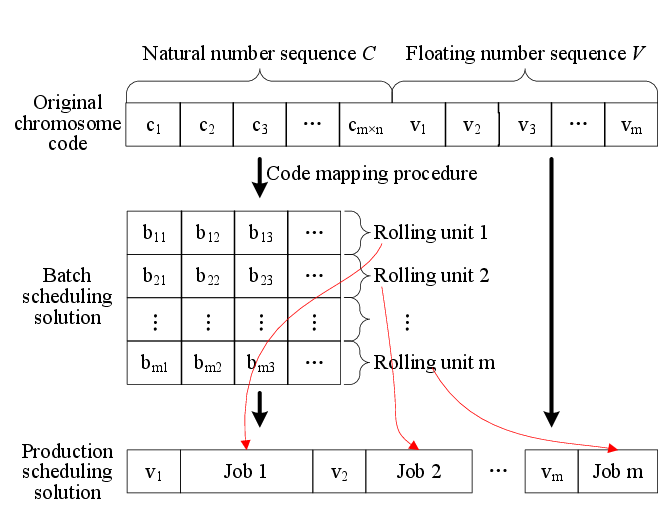}
\caption{Relationship between production scheduling and electricity costs under TOU pricing}
\label{fig:4}       % Give a unique label
\end{center}
\end{figure}
In order to contain information both of batch scheduling and job-shop scheduling, a hybrid chromosome code consists of two sections as shown in Figure~\ref{fig:4} is designed. The first section is a natural number sequence $C$ that can be transformed to a two-dimensional matrix $B$ through a code mapping procedure, where $B$ represents a batch scheduling solution and element $b_{ij}$ in $B$ is the original sequence of slab $j$ in rolling unit $i$. For each $i$, if the minimal $j$ is found while $b_{ij}=0$, it can be resolved that the last slab in rolling unit $i$ is $b_{i,j-1}$. The second section is a floating number sequence $V$ that represents the idle time allocated during job-shop scheduling, where job means production of rolling units.

According to above description, the hybrid chromosome code $G$ can be expressed as
\begin{equation}\label{eq:3-1}
  \left\{ \begin{array}{l}
  G=(C,V)\\
  C=(c_1,c_2,\ldots,c_{m\times n})\\
  V=(v_1,v_2,\ldots,v_m)
  \end{array} \right.,
  \nonumber
\end{equation}
where element $c_i$ in $C$ is a natural number that ranged from 1 to $m\times n$, $m$ is the quantity of rolling units and $n$ is the quantity of slabs to be scheduled, any two number $c_i$ and $c_j$ are assigned to different values, $v_i$ in $V$ represents the idle time allocated to rolling unit $i$ before rolling production.

Detailed steps of the code mapping procedure as mentioned previously are listed as follows:

\paragraph{Step 1} Set $f_i(i=1,2,\ldots,n)$ to 0, where $f_i$ is a flag and $f_i=1$ represent slab $i$ has been scheduled into rolling units; for rolling unit $k$ ($k=1,2,\ldots,m$), set $num_k=0$, where $num_k$ means the slab quantity in rolling unit $k$; set $d_k=0$, where $d_k$ is the accumulative rolling length in rolling unit $k$, set $q_k=0$, where $q_k$ means the continuously rolled length of slabs with same width in rolling unit $k$; define a loop variable $j$ and set $j=1$;
\paragraph{Step 2} Confirm the variables $s$ and $k$ in accordance with natural number $c_j$, by which slab $s$ is scheduled in rolling unit $k$ can be determined. $s$ and $k$ can be calculated by
\begin{equation}\label{eq:calc_s}
  s=c_j-\left[\frac{c_j-1}{m}\right]\times m
  \nonumber
\end{equation}
and
\begin{equation}\label{eq:calc_k}
  k=\left[\frac{c_j-1}{n}\right]+1.
  \nonumber
\end{equation}
\paragraph{Step 3} Check if condition $f_s=0$ is satisfied:\\
i. If satisfied, it means that slab $s$ is an unscheduled slab. Then if $w_s\neq w_k'$, set $q_k=0$, where $w_s$ is the width of the slab $s$ and $w_k'$ is the width of the latest appended slab in rolling unit $k$. Furthermore, if $d_k+l_s\leq U$ and $q_k+l_s\leq R$, put slab $s$ into rolling unit $k$ and update matrix $B(=[b_{ij}])$ by $b_{k,num_k}=s$, set $num_k=num_k+1$, $d_k=d_k+l_s$, $q_k=q_k+l_s$ and $f_s=1$;\\
ii. Otherwise, go to step 4;
\paragraph{Step 4} Update $j=j+1$, go to step 2 to repeat the above operations until $j=m\times n+1$;
\paragraph{Step 5} Check if $f_i=1(i=1,2,\ldots,n)$ and $d_k\geq L(k=1,2,\ldots,m)$ are all satisfied:\\
i. If satisfied, it means that all slabs are scheduled into rolling units with subjection to given constraints. Perform idle time allocation procedure to generate sequence $V$ of chromosome code, then $G=(C, V)$ represent a feasible solution of the model in this paper;\\
ii. Otherwise, a large number should be assigned to function $f_1$ and $f_2$ to avoid chromosome be selected into new population in next selection operator.

\parskip=1\baselineskip
The detailed steps of idle time allocation as mentioned above based on section $C$ of chromosome code are listed as follows:
\parskip=0\baselineskip
\paragraph{Step 1} Confirm the electricity price $\pi_i^s$ corresponding to each rolling unit $i$ when the production start and price $\pi_i^e$ when the production complete;
\paragraph{Step 2} Create a random floating number sequence $V$ that represents the idle time allocated for rolling units before production. For elements in $V$, the constraint as Eq. (\ref{eq:2-8}) in Section \ref{sec:2.2} must be satisfied;
\paragraph{Step 3} Sort time periods in descending order based on electricity price, after that a new set of time periods $T'=(t_1',t_2',\ldots,t_t')$ is generated, in which the price associated with $t_k'$ is $\pi_k'$; define a loop variable $j$ and set $j=1$;
\paragraph{Step 4} Adjust the idle time allocation for rolling units. For each rolling unit $i$ that started from time period $t_j'$, if $\pi_i^e<\pi_j'$ and $v_{i+1}>0$, set $v_{i+1}=0$, $v_i=v_i+v_{i+1}$; for rolling unit $i$ that completed in time period $t_j'$, if $\pi_i^s<\pi_j'$ and $v_i>0$, set $v_i=0$, $v_{i+1}=v_{i+1}+v_i$;

\paragraph{Step 5} Update variable $j=j+1$, go to step 4 to repeat the above operation for the left time periods until $j=t$, which represent adjustment of idle time allocation is completed.

\parskip=1\baselineskip
The benefits of hybrid encoding and mapping procedure above are not only containing complete information of production scheduling but also handling constraints. From step 5, we can see that all constraints from Eq. (\ref{eq:2-3})--(\ref{eq:2-7}) in Section \ref{sec:2.2} are satisfied in accepted feasible solution, which is helpful to reduce the difficulty of problem solving.
\parskip=0\baselineskip

\subsubsection{Design of genetic operators}
\label{sec:3.2}

In order to instantiate the MOPSA algorithm, customized genetic operators are defined to match hybrid chromosome code, the most important operators for genetic algorithm are selection, crossover, and mutation.

Selection operator, which means selecting individuals from population, is done based on the frontier rank of individuals by non-dominated sorting. If many individuals have the same rank, the individual with maximum crowded distance will be selected preferentially.

Partially Mapped Crossover (PMX) that mentioned in \cite{tasan2012genetic} and Scramble Sub-list Mutation (SSM) mentioned in \cite{mutlu2013iterative} are adopted to perform operation on section $C$ of chromosome code. The PMX operator is performed on two parent chromosomes: randomly select two crossover points $k_1$ and $k_2$ and separate the chromosome code into three sections, swap the gene codes in range $[k_1, k_2]$, after that, replace the other gene codes out of range $[k_1, k_2]$ according to mapping relationship that determined by the middle section.

Unlike the crossover operator, SSM mutation operator is performed on single parent chromosome: randomly select two positions $p_1$ and $p_2$ that separated less than a fixed length in the chromosome code, then rearrange the gene codes between $[p_1, p_2]$.

After crossover or mutation operation, update section $V$ of the chromosome code to allocate idle time for rolling units immediately.

\subsubsection{Decision expressions and fitness function calculation}
\label{sec:3.3}
We choose the objective functions $f_1$ and $f_2$ to be the fitness functions in our genetic algorithm. $f_1$ represents penalties and $f_2$ represents electricity costs in production, which are both cost-oriented that need to find minimum value.

In fitness function calculation, most needed variables and expressions are static and can be pre-computed except the variable $x_{ij}^k$ in $f_1$ and $d_i^j$ in $f_2$, so the key of fitness function calculation is to determine the value of $x_{ij}^k$ and $d_i^j$ based on chromosome code.

According to the characteristics of chromosome code in this paper, we use matrix $B(=[b_{ij}])$ that generated in chromosome code mapping procedure instead of part $C$ to perform the following calculation.
In order to determine the value of $x_{ij}^{k}$, each row in matrix $B$ should be traversed to search the adjacent elements that satisfy the following equation
\begin{equation}\label{eq:2-15}
\left\{ \begin{array}{ll}
b_{k,j_1}=i, \\
b_{k,j_1+1}=j
\end{array} \right.
\end{equation}
where the first equation means slab $i$ is assigned in rolling units $k$ and processed with the sequence $j_1$ , and the next equation indicates that slab $j$ is allocated after slab $i$ immediately in rolling unit $k$. $x_{ij}^{k}$ can be determined to be 1 if Eq. (\ref{eq:2-15}) is satisfied, otherwise 0. For each rolling units $k$, penalties between adjacent slabs are accumulated by $P_{ij}\cdot x_{ij}^k$;

Meanwhile, it should be noted that calculation of $d_{i}^{j}$ in $f_2$ would not only depend on $B$ but also on sequence $V$ that represents the allocated idle time for rolling units. According to Eq. (\ref{eq:2-13})--(\ref{eq:2-14}) that defined in Section \ref{sec:2.2}, The determination of $d_{i}^{j}$ mainly depend on variables $y_i^k$, $r_{ij}^k$ and $v_i$, in which the first two variables can be easily calculated on matrix $B$ by a traversal procedure as done in determining $x_{ij}^{k}$, and the last variable $v_i$ can be directly identified by the sequence $V$ in chromosome code.
Once $d_i^j$ is known, fitness function $f_2$ can be accumulated by $\pi_j\cdot W_i\cdot d_i^j$ for each time periods.

\subsection{TOPSIS based multi-criteria decision-making}
\label{sec:3.2}
As MOPSA generate more than one Pareto-optimal solution, in order to facilitate field operation, only a few solutions should be accepted. In this paper, Technique for Order Preference by Similarity to an Ideal Solution (TOPSIS) \cite{tzeng2011multiple}, a widely used multi-criteria decision-making method to identify solutions from finite alternatives, is adopted as the method to select a recommended optimal solution.

Detailed steps of the TOPSIS based multi-criteria decision-making for HRPSP are listed as follows:
\paragraph{Step 1:} The decision matrix $X$ can be expressed as
\begin{equation}\label{eq:3-3}
  X=
    \left[ \begin{array}{cc}
    x_{11} & x_{12}\\
    x_{21} & x_{22}\\
    \vdots & \vdots\\
    x_{m1} & x_{m2}
    \end{array} \right],
    \nonumber
\end{equation}
where $X$ is a two dimensional matrix with the size of $m\times n$, which means that there're $m$ solutions generated by the multi-objective algorithm and $n$ objectives for the HRPSP, where $n=2$. The element $x_{ij}$ in $X$ is the value of the $j$th objective with respect to the $i$th solution.
Then the normalized decision matrix $Z(=[z_{ij}])$ can be calculated according to
\begin{equation}\label{eq:3-4}
    z_{ij}=\frac{x_{ij}}{\sqrt{\sum_{i=1}^mx_{ij}^2}}.
    \nonumber
\end{equation}
\paragraph{Step 2:}Multiply the normalized decision matrix by its associated weights to calculate the weighted normalized decision matrix $V(=[v_{ij}])$, in which $v_{ij}$ is calculated as
\begin{equation}\label{eq:3-5}
    v_{ij}=w_j\cdot z_{ij},
    \nonumber
\end{equation}
where $w_j$ is a weight factor associated with the $j$th objective. In our context, $w_1$ and $w_2$ are set to different values according to preference of two objectives.
\paragraph{Step 3:}Identify the the ideal solution $s^+$ and the nadir solution $s^-$ of each objective according to the following equations:
\begin{equation}\label{eq:3-6}
    s^+=\left(s_1^+,s_2^+\right),
    \nonumber
\end{equation}
\begin{equation}\label{eq:3-7}
    s_j^+=\left\{ \begin{array}{ll}
    \max\limits_{1\leq i\leq m}v_{ij} & \textrm{if $f_j$ is benefit-oriented,} \\
    \min\limits_{1\leq i\leq m}v_{ij} & \textrm{if $f_j$ is cost-oriented.}
    \end{array} \right.
    \nonumber
\end{equation}
\begin{equation}\label{eq:3-8}
    s^-=\left(s_1^-,s_2^-\right),
    \nonumber
\end{equation}
\begin{equation}\label{eq:3-9}
    s_j^-=\left\{ \begin{array}{ll}
    \min\limits_{1\leq i\leq m}v_{ij} & \textrm{if $f_j$ is benefit-oriented,} \\
    \max\limits_{1\leq i\leq m}v_{ij} & \textrm{if $f_j$ is cost-oriented.}
    \end{array} \right.
    \nonumber
\end{equation}

It should be known that both of the objectives in HRPSP are cost-oriented, which is said to find the minimum of objective functions.
\paragraph{Step 4:}Measure the distances $d_i^+$ and $d_i^-$ of the $i$th solution from the ideal solution $s^+$ and the nadir solution $s^-$ by
\begin{equation}\label{eq:3-10}
    d_i^+=\sqrt{\sum_{j=1}^n\left(v_{ij}-s_j^+\right)^2}, i=1,2,\ldots,m,
    \nonumber
\end{equation}
\begin{equation}\label{eq:3-11}
    d_i^-=\sqrt{\sum_{j=1}^n\left(v_{ij}-s_j^-\right)^2}, i=1,2,\ldots,m.
    \nonumber
\end{equation}
\paragraph{Step 5:}Calculate $C_i^*$ that represents the relative closeness of $i$th solution with respect to the ideal solution according to
\begin{equation}\label{eq:3-12}
    C_i^*=\frac{d_i^-}{\left(d_i^-+d_i^+\right)}, i=1,2,\ldots,m.
    \nonumber
\end{equation}

\parskip=0\baselineskip
After completing the above steps, the decision-making can be finally performed on the Pareto-optimal solutions according to the sequence that determined by  $C_i^*(i=1,2,\ldots,m)$ in descending order, the solution that owns maximal relative closeness will be selected as the recommended optimal solution.
\parskip=0\baselineskip

\section{Experimental results and analyses}
\label{sec:4}
In this section, we perform a series of experiments to evaluate the effectiveness and performance of the proposed method in different scenario.
\subsection{Experimental procedure}
\label{sec:4.1}
In experimental procedure, four groups of production data as is shown in Table \ref{tab:1} are collected from a steel mill for experimental procedure. For each group of production data, if there are many slab varieties in width, gauge and hardness, the penalty score between adjacent slabs will be larger. At the same time, full production load means the idle time for processing slabs will be short.

According to constraints of production equipment and capability, the lower and upper bound of the cumulative length of slabs that scheduled in a single rolling unit are respectively set to 5 and 10 kilometer, and the upper bound of the continuously rolled length of slabs with same width is set to 1 kilometer. For specific slab, rolling length, processing time and power consumption can be obtained by the hot rolling process control system in steel mill. The penalties that caused by jumps between adjacent slabs in width, gauge and hardness are adopted by referencing to \cite{kosiba1992discrete}.
The data in Table \ref{tab:2} is an actually performed TOU electricity tariffs in steel mill. According to daily power load distribution, a whole day is split into eight periods that contain four types of time periods, each type of time period is associated with corresponding price.

\begin{table*}[btp]
\caption{Production data description}
\label{tab:1}
\begin{tabular}{lllll}
\hline\noalign{\smallskip}
\tabincell{l}{Group Id} & \tabincell{l}{Slab \\quantity} & \tabincell{l}{Rolling units \\quantity} & \tabincell{l}{Processing \\time /(Min)} & Characteristics\\
\noalign{\smallskip}\hline\noalign{\smallskip}
1 & 450 & 8 & 1421.75 & Many varieties of slabs and full production load\\
2 & 415 & 8 & 1323.05 & Many varieties of slabs and not full production load\\
3 & 450 & 8 & 1427.88 & Few varieties of slabs and full production load\\
4 & 415 & 8 & 1318.33 & Few varieties of slabs and not full production load\\
\noalign{\smallskip}\hline
\end{tabular}
\end{table*}

\begin{table*}[btp]
\caption{TOU electricity tariffs}
\label{tab:2}
\begin{tabular}{lll}
\hline\noalign{\smallskip}
\tabincell{l}{Time period} & Time frame & \tabincell{l}{Electricity price /(CNY$\cdot$kWh$^{-1}$)}\\
\noalign{\smallskip}\hline\noalign{\smallskip}
on-peak & 18:00-21:00 & 0.878\\
mid-peak & 08:00-11:00, 15:00-18:00 & 0.778\\
flat-peak & \tabincell{l}{07:00-08:00, 11:00-15:00, 21:00-22:00} & 0.628\\
off-peak & 00:00-07:00, 22:00-24:00 & 0.428\\
\noalign{\smallskip}\hline
\end{tabular}
\end{table*}

\begin{table*}[tp]
\caption{Scheduling results obtained by different methods}
\label{tab:3}
\begin{tabular}{p{3cm}p{0.8cm}p{1.5cm}p{0.8cm}p{1.5cm}p{0.8cm}p{1.5cm}p{0.8cm}p{1.5cm}}
\hline
\multirow{2}{*}{\tabincell{l}{Method}} & \multicolumn{2}{l}{Group 1} & \multicolumn{2}{l}{Group 2} & \multicolumn{2}{l}{Group 3} & \multicolumn{2}{l}{Group 4}\\
\cline{2-9}
 & $f_1$ & $f_2$ & $f_1$ & $f_2$ & $f_1$ & $f_2$ & $f_1$ & $f_2$ \\
\hline
CM1                     & 5035 & 313254 & 4528 & 296357 & 2957 & 315894 & 2659 & 299623 \\
CM2                     & 5035 & 309078 & 4528 & 276813 & 2957 & 312753 & 2659 & 278717 \\
PM, $w=[0.9, 0.1]$      & 5129 & 308281 & 4573 & 275898 & 3090 & 311114 & 2710 & 277214 \\
PM, $w=[0.4, 0.6]$      & 7493 & 305691 & 6905 & 274022 & 3445 & 309242 & 3022 & 275397 \\
PM, $w=[0.1, 0.9]$      & 7701 & 305680 & 7665 & 273729 & 3478 & 309234 & 3308 & 274994 \\
\hline
\end{tabular}
\end{table*}

In order to obtain excellent algorithm performance, the NSGA-II parameters are determined by parameter sensitivity analysis based on empirical value and a lot of tests. The probability of crossover and mutation are set to 0.4 and 0.6 respectively, the population size is set to 50, the maximum iterations of algorithm is set to 5000. The production scheduling optimization algorithm and TOPSIS decision making procedure are both implemented and performed in MATLAB.

In experimental procedure, the proposed method (named as PM) are compared with two conventional methods to evaluate effectiveness and performance. Since exact algorithm for large scale HRBSP problem is too difficult to implement, genetic algorithm is often used for solving this problem. In this paper, a relatively new method in reference \cite{chen2012development} with the traditional objective to minimize jump penalties is adopted as a comparison method (named as CM1), in which a hybrid evolutionary algorithm with integration of genetic algorithm and extremal optimization is designed to solve the hot rolling scheduling problem.

Because electricity price during hot rolling change over time, it is natural to allocate the processing sequence and the idle time of rolling units to avoid on-peak time periods, then the MILP method proposed by \cite{wang2012integrated} is adopted as a another comparison method (named as CM2) to find the low bound of electricity costs on the basis of solution obtained in CM1.

Unlike single objective optimization, the results of multi-objective optimization is not a single solution but a set of Pareto-optimal solutions, in order to facilitate field operation, we choose different values of objective weight factors $w_j$ in TOPSIS decision-making procedure to recommend solution with different preference of penalty score and electricity cost. In our experimental procedure, the objective factors $w(=[w_1,w_2])$ of the proposed method are set to [0.9,0.1], [0.4,0.6] and [0.1,0.9] respectively.

\begin{table*}[tbp]
\begin{threeparttable}
\caption{Detailed parameters of scheduling results for group 1 of production data}
\label{tab:4}
\begin{tabular}{llp{0.8cm}p{0.8cm}p{0.8cm}p{1.2cm}p{1.2cm}p{1.2cm}p{1.2cm}p{1.2cm}}
\hline\noalign{\smallskip}
\tabincell{l}{Method} & \tabincell{l}{RUS} & \tabincell{l}{SQ} & \tabincell{l}{RL} & \tabincell{l}{PT} & \tabincell{l}{PD} & \tabincell{l}{APL} & \tabincell{l}{PST} & \tabincell{l}{PCT} & \tabincell{l}{AIT}\\
\noalign{\smallskip}\hline\noalign{\smallskip}
\multirow {8}{*}{\tabincell{l}{PM, $w=[0.1, 0.9]$}}
& 1	& 51	& 8.53	& 2.44	& 56.74	& 23.25	& 00:00	& 02:26	& 0\\
& 2	& 58	& 9.95	& 2.79	& 64.17	& 23.00	& 02:26	& 05:14	& 0\\
& 3	& 55	& 9.92	& 2.80	& 60.75	& 21.70	& 05:14	& 08:02	& 0\\
& 4	& 61	& 9.94	& 3.44	& 68.20	& 19.83	& 08:02	& 11:28	& 0\\
& 5	& 59	& 9.96	& 3.16	& 66.46	& 21.03	& 11:28	& 14:38	& 0\\
& 6	& 57	& 9.81	& 3.17	& 62.24	& 19.63	& 14:38	& 17:48	& 0\\
& 7	& 52	& 9.10	& 2.89	& 54.39	& 18.82	& 17:48	& 20:41	& 0\\
& 8	& 57	& 9.94	& 3.00	& 66.69	& 22.23	& 20:59	& 23:59	& 0.3\\
\hline\noalign{\smallskip}
\multirow {8}{*}{\tabincell{l}{CM1}}
& 1	& 57	& 9.80 	& 3.07	& 63.38	& 20.64	& 00:00	& 03:04	& 0  \\
& 2	& 62	& 10.00	& 3.23	& 67.76	& 20.98	& 03:04	& 06:18	& 0  \\
& 3	& 56	& 9.92 	& 2.96	& 63.13	& 21.33	& 06:18	& 09:16	& 0  \\
& 4	& 55	& 9.47 	& 2.96	& 62.15	& 21.00	& 09:16	& 12:14	& 0  \\
& 5	& 58	& 9.91 	& 2.99	& 63.72	& 21.31	& 12:14	& 15:13	& 0  \\
& 6	& 53	& 9.53 	& 2.83	& 59.68	& 21.09	& 15:13	& 18:03	& 0  \\
& 7	& 51	& 8.54 	& 2.70	& 56.86	& 21.06	& 18:03	& 20:45	& 0  \\
& 8	& 58	& 9.98 	& 2.96	& 62.95	& 21.27	& 20:45	& 23:42	& 0  \\

\noalign{\smallskip}\hline
\end{tabular}
\small
\begin{tablenotes}
  \item Abbreviation: RUS--rolling unit sequence, SQ--slab quantity, RL--rolling length (km), PT--processing time (h), PD--power demand (MW$\cdot$ h), APL--average power load (MW), PST--processing start time (HH:mm), PCT--processing complete time (HH:mm), AIT--allocated idle time (Hour).
\end{tablenotes}
\end{threeparttable}
\end{table*}

\begin{figure*}[tbp]
\centering
%\begin{center}
\subfigure[group 1]{\includegraphics[width=0.48\textwidth,height = 5.8cm]{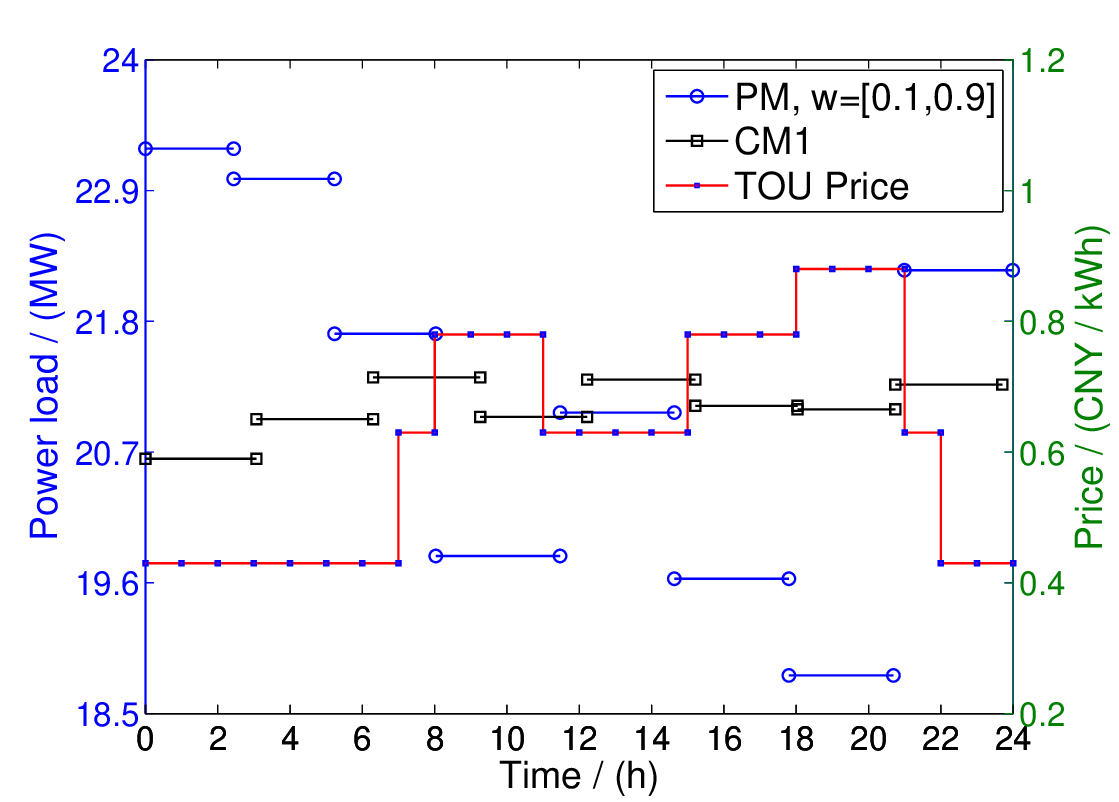}}
\subfigure[group 2]{\includegraphics[width=0.48\textwidth,height = 5.8cm]{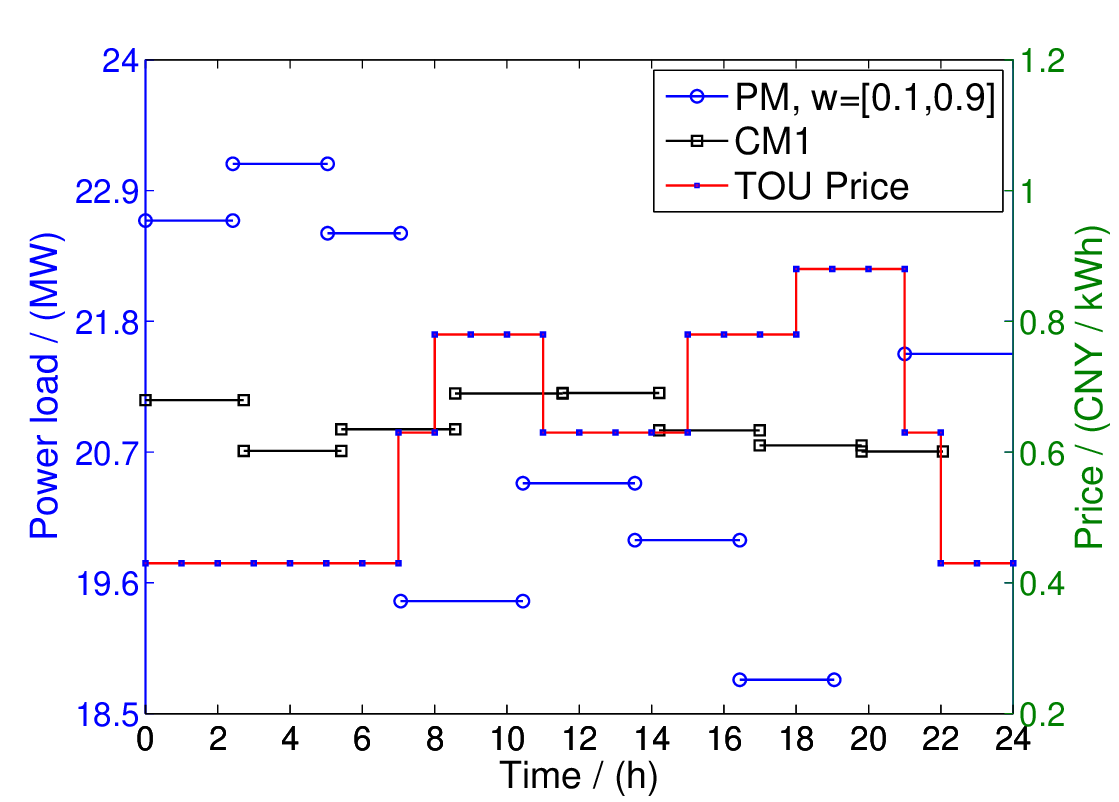}}
\subfigure[group 3]{\includegraphics[width=0.48\textwidth,height = 5.8cm]{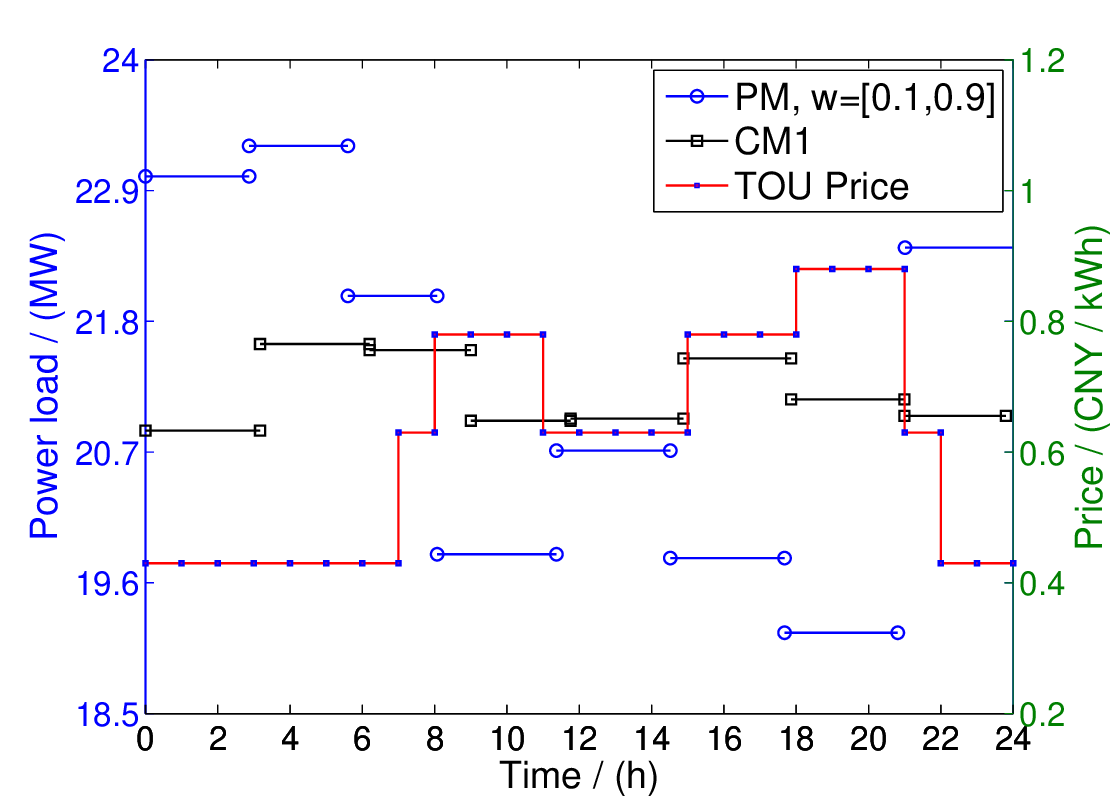}}
\subfigure[group 4]{\includegraphics[width=0.48\textwidth,height = 5.8cm]{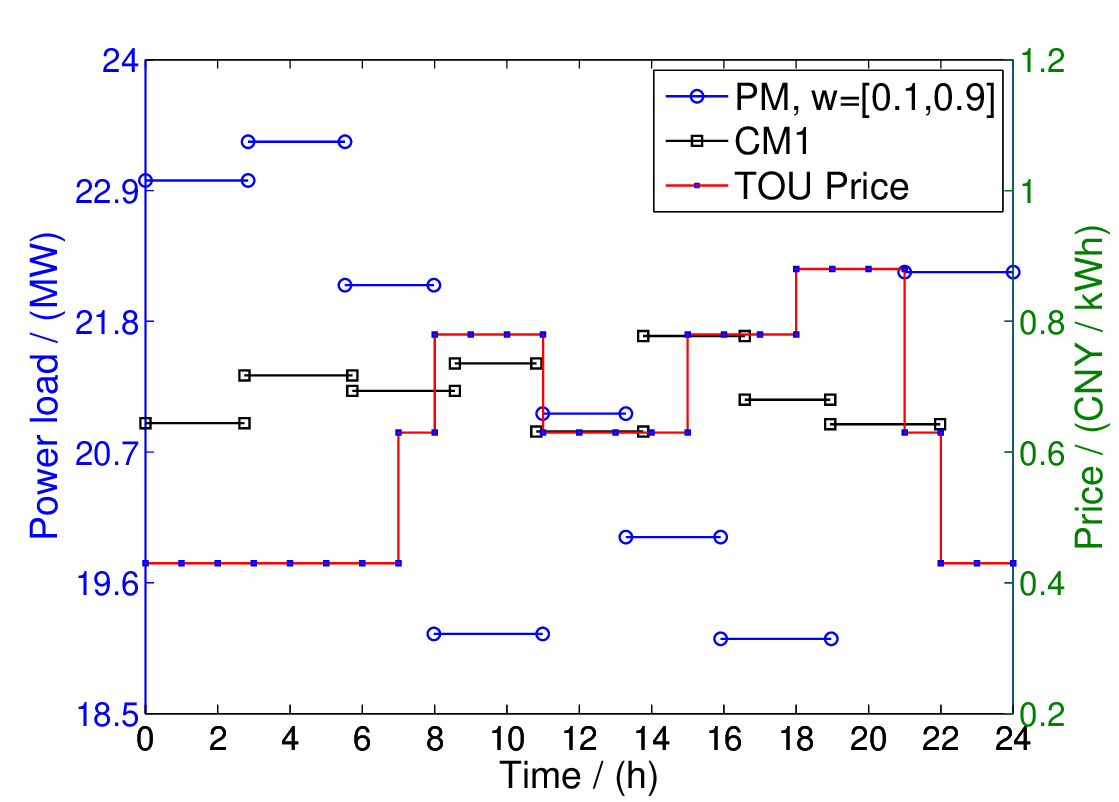}}
\caption{Illustration of job scheduling results obtained by PM and CM1}
\label{fig:5}
%\end{center}
\end{figure*}

\begin{figure*}[tbp]
\begin{center}
\subfigure[group 1]{\includegraphics[width=0.48\textwidth]{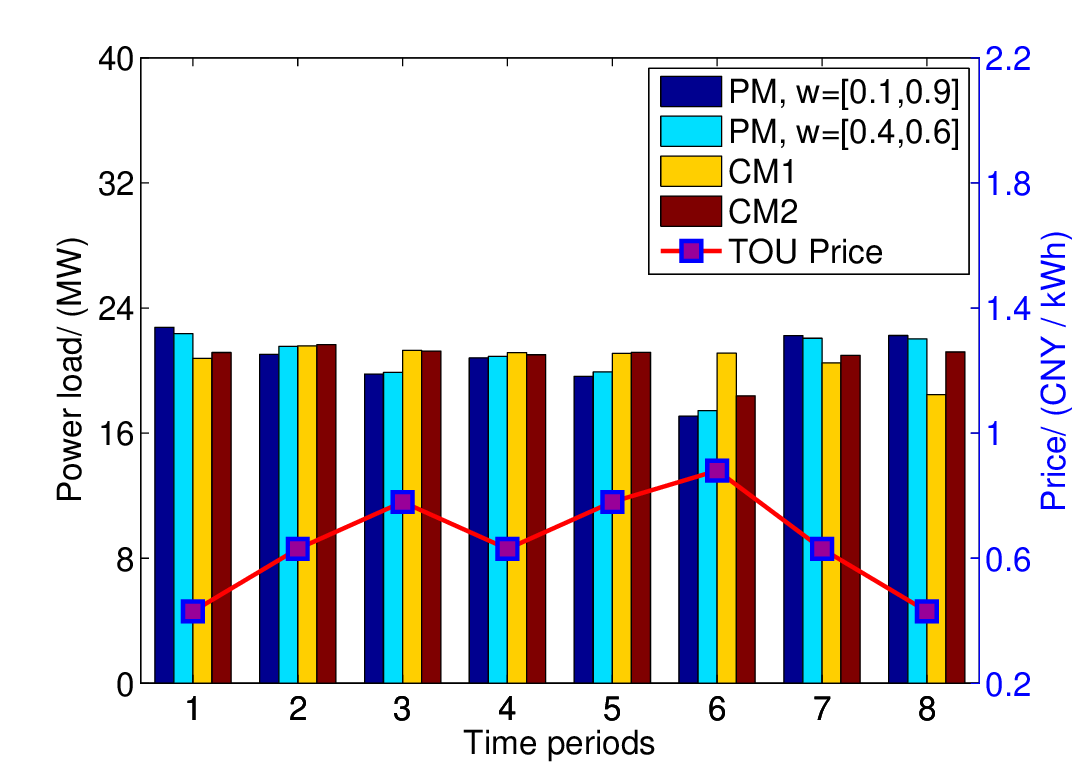}}
\subfigure[group 2]{\includegraphics[width=0.48\textwidth]{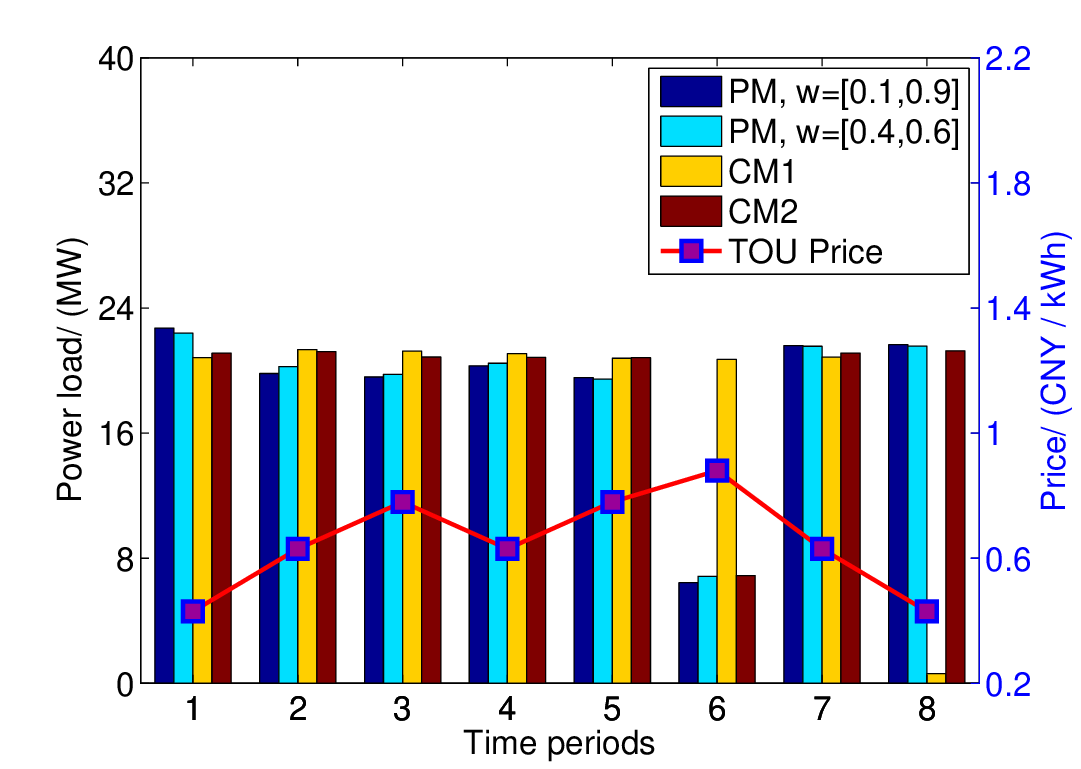}}
\subfigure[group 3]{\includegraphics[width=0.48\textwidth]{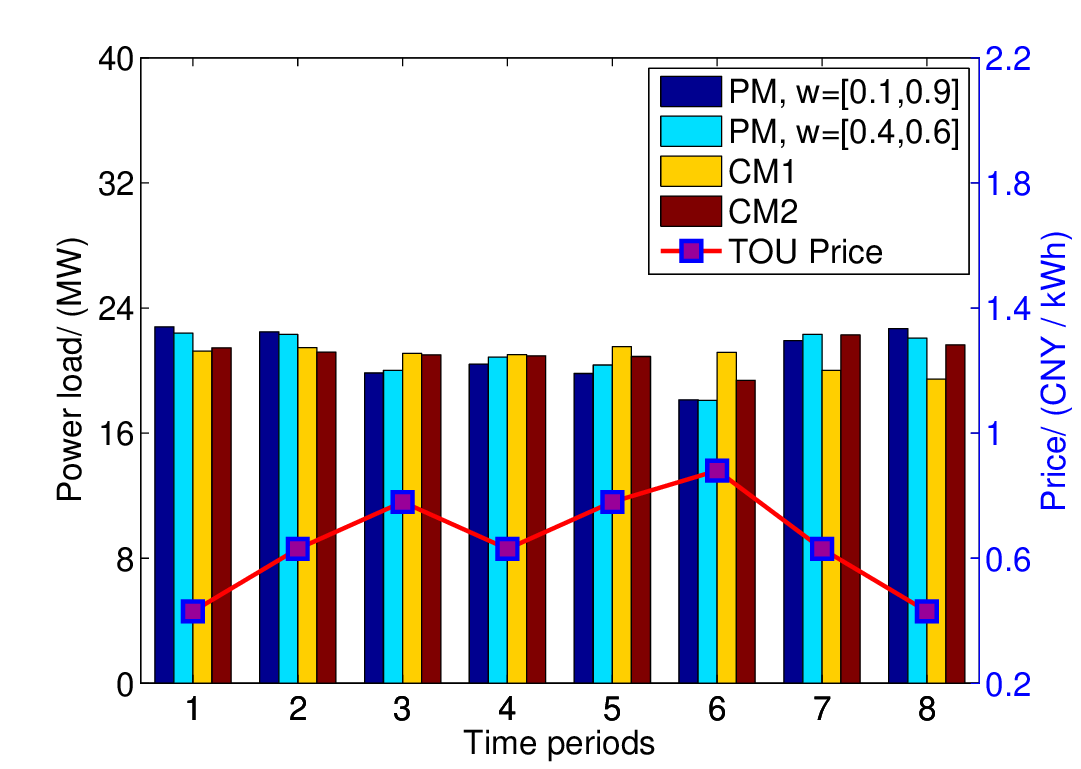}}
\subfigure[group 4]{\includegraphics[width=0.48\textwidth]{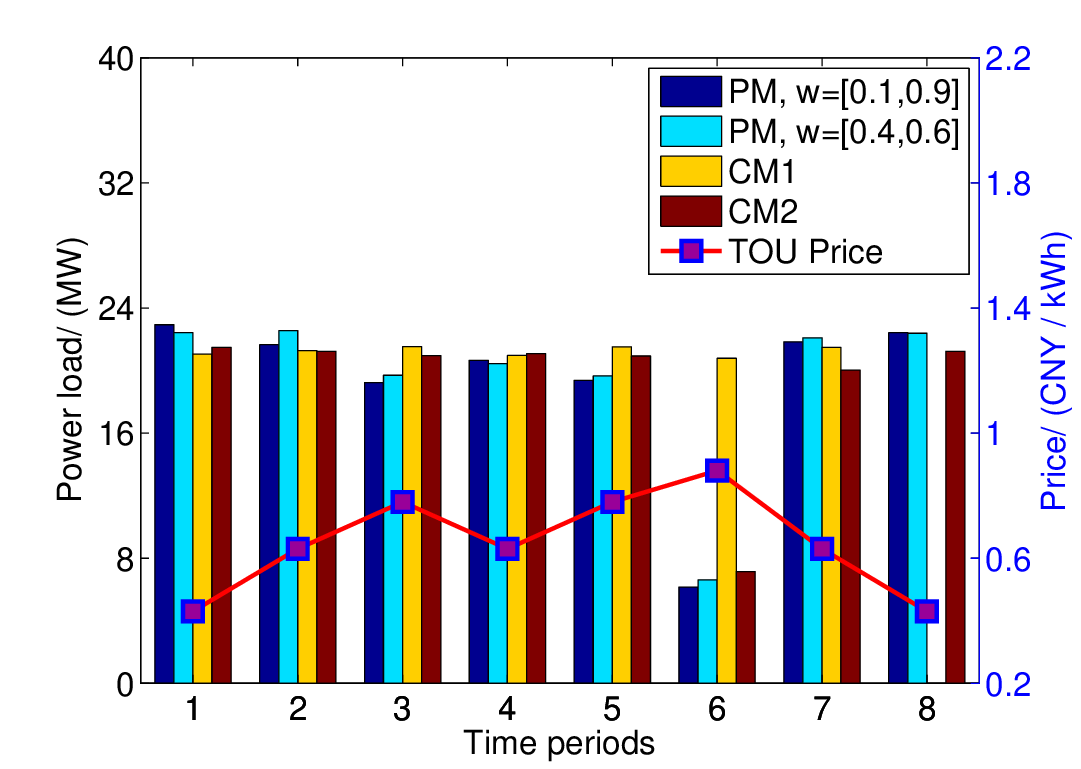}}
\caption{Power load distribution among time periods}
\label{fig:6}
\end{center}
\end{figure*}

Optimization results obtained by different methods are provided in Table \ref{tab:3}, in which we can see that penalties obtained by PM with $w=[0.1,0.9]$ are roughly the same as that obtained by CM1 and CM2 but electricity costs cut down obviously. It is obvious that load shifting to reduce electricity cost inevitably result in an increasing of penalty score, and we'd just like to point out that minimizing jump penalties is a guiding target but not a strictly rigid constraint in engineering. If there is allowance for penalty increase on electricity cost, more significant effect on electricity cost reduction is shown, which tell us that penalty relaxation can play an import role while electricity costs is the key consideration in production, as a consequence, we can utilize objective weight factors in TOPSIS procedure to adjust preferences of the two objectives. In our cases, electricity cost obtained by PM with TOPSIS decision making on each group of data is less than CM1, even compared to CM2, which includes load shifting on fixed rolling batches, the result is still better, this advantage is attributed to TOU pricing based batching to construct rolling units.
Besides that, we can see that the optimization effect is more significant while the production load is not full, i.e., group 2 and 4, which is caused by more idle time margin existed to avoid on-peak time periods in such situation.

\subsection{Scheduling results analysis}
\label{sec:4.2}
In this section, group 1 of data is chosen to have a detailed analysis on job scheduling results firstly. Because the main idea of this paper is ELD, the proposed method PM with $[w_1,w_2]=[0.1,0.9]$, which has the most significant effect on electricity cost reduction, is selected to compare with the conventional method CM1. Rolling parameters obtained by both methods are given in Table \ref{tab:4}, from which we can see that the parameters subject to instantiated constraints, which represents that the schedule is feasible solution. Then, we analyze the scheduling results from two aspects.

On one hand, rolling units in Table \ref{tab:4} is considered as production jobs and illustrated in Figure~\ref{fig:5}. As it can be seen, in any sub figure, heavy loads are allocated in off-peak and flat-peak periods by PM, while light loads are allocated in on-peak or mid-peak periods. In addition, idle time is allocated at 18:00 to 21:00 for our scenarios. Another phenomenon is that the power load difference between heavy load and light load in PM is greater than that in CM1 and CM2, which is due to that rolling units in PM are organized by TOU electricity price and their processing time.

On the other hand, average power load distribution among time periods are illustrated in Figure~\ref{fig:6}.
Compared to CM1, power load obtained by PM reduce greatly in the last on-peak periods and increase substantially in last off-peak period, especially for group 2 and group 4, which are characterized by not full production load. At the same time, power load in the first off-peak period increase in a certain extent. In addition, power load distribution obtained by PM is also better than that obtained by CM2 that based on the principle of load shifting correspond to TOU pricing, which confirm the effectiveness and advancement of the proposed method furthermore.

From above results and analyses, we know that the advantages of our proposed method on electricity cost reduction can be attributed to two aspects, one is load shifting to avoid on-peak time periods, and the other one is TOU pricing based load planning.

\subsection{Robustness of the algorithm}
\label{sec:4.3}
It is well known that NSGA-II is a randomized algorithm, each run of the algorithm may get different results. For evaluating robustness of the algorithm, we use box plot to portray the convergence metric in repeated operation, which is represented by average value of minimum normalized Euclidean distance and indicates the disparity between approximate Pareto-frontier and ideal Pareto-frontier.

Assume that $P^*=(p_1,p_2,\ldots,p_{|P^*|})$ is the optimal solutions that evenly distributed on ideal Pareto-frontier, and $A^*=(a_1,a_2,\ldots,a_{|A^*|})$ is the approximate solutions obtained in a single run of the proposed algorithm. For any $a_i$, minimum normalized Euclidean distance $d_i$ between $a_i$ and $P^*$ can be calculated by
\begin{displaymath}
    d_i=\min_{j=1}^{|P^*|} \sqrt {\sum_{m=1}^{2} \left(\frac{f_m(a_i)-f_m(a_j)}{f_m^{max}-f_m^{min}}\right)^2},
\end{displaymath}
where $f_m^{max}$ and $f_m^{min}$ respectively represent the maximum and minimum value of the $m$th objective function in $P^*$, and then the convergence metric $C$ can be expressed as
\begin{displaymath}
    C(A^*)\triangleq \frac{\sum_{i=1}^{|A^*|}d_i}{|A^*|}.
\end{displaymath}

\begin{figure}[tbp]
\begin{center}
\subfigure{\includegraphics[width=0.5\textwidth]{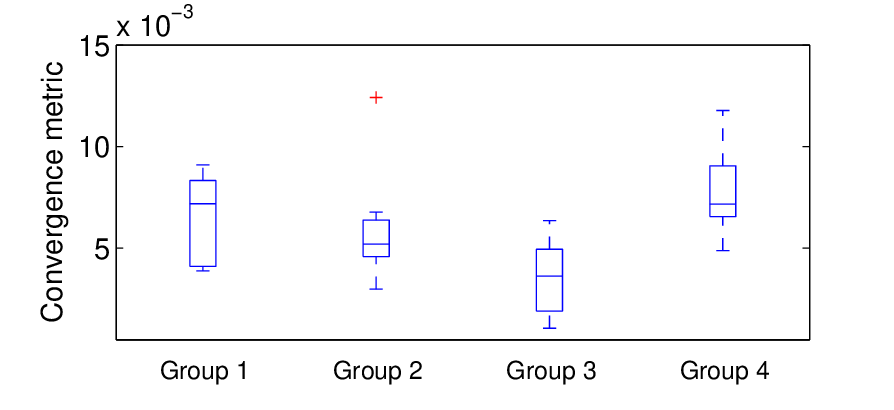}}
\caption{Box plots based on convergence metrics}
\label{fig:7}
\end{center}
\end{figure}

Note that the ideal Pareto-frontier are always unknown in real problem, the algorithm proposed in this application are run 30 runs respectively on each group of production data, and then a pseudo Pareto-frontier, which consist of all the solutions in 30 times run with removing dominated solutions, is constructed to compare with the approximate Pareto-frontiers.
For every run, box plots based on convergence metrics are illustrated in Figure~\ref{fig:7}. In general, metric $C$ in less than $10^{-2}$ means good statistical convergence performance in Pareto optimality based multi-objective optimization. The symbol ``+" in Figure 7 refers to an outlier in box statistics; nevertheless, it can be seen that the outlier is very close to $10^{-2}$. Overall, we can see that the upper edges on different groups of data are all less than $10^{-2}$, except a slightly larger value on group 4 and an outlier on group 2. Even so, the 3rd quartile on group 4 is totally in the range of less than $10^{-2}$. The statistical results show that the proposed algorithm is stable in repeated run. On the whole, we can conclude that the proposed algorithm is robust and suitable for application in engineering.

\section{Conclusions}
\label{sec:5}
This paper presented the challenge of energy saving in hot rolling production and formulated a multi-objective optimization model of HRPSP under TOU electricity pricing. Objective of the model is to minimize electricity costs in production while considering penalties caused by jumps between adjacent slabs.
Since exact algorithm is difficult to implement for solving the large scale problem, a NSGA-II based production scheduling algorithm was developed to obtain Pareto-optimal solutions, and then TOPSIS decision making method was adopted to recommend solution with different objective preferences. Experimental results and analyses showed that the proposed method cut electricity costs in production, and the performance is better than load shifting
on fixed production load.
Consider multiple production lines existed in most steel mills, HRPSP integrated multiple parallel machine job-shop scheduling will be the subject of further study, which is expected to have greater benefits. Besides that,multistage scheduling problem will also be our next work.

%\begin{acknowledgements}
%This work was supported by National Natural Science Foundation of China (61402391, 61573299) and Natural Science Foundation of Hunan Province of China (2015JJ5027).
%\end{acknowledgements}

%\bibliography{mybib2}   % name your BibTeX data base
%\bibliographystyle{spphys}


\bibliographystyle{spphys}

\end{document}